\documentclass[a4paper,USenglish]{lipics-wst2018}

\usepackage{xcolor}
\usepackage{xspace}
\usepackage{enumitem}
\usepackage{fontawesome}
\usepackage{rail}

\renewenvironment{quote}
  {\small\list{}{\rightmargin=0cm \leftmargin=\leftmargin}%
   \item\relax}
  {\endlist}

\newcommand{\arxiv}[1]{
  \href{http://arxiv.org/abs/#1}{\nolinkurl{arXiv:#1}}}

\newcommand\toolname[1]{\mbox{\textsf{#1}}}
\newcommand\tool[1]{\toolname{#1}\xspace}
\newcommand\aprove{\tool{APro\kern-0.1exVE}}
\newcommand\ttt[1]{\tool{%
  T\kern-0.15em\raisebox{-0.55ex}T\kern-0.15emT\kern-0.15em\raisebox{-0.55ex}{#1}}}
\newcommand\ceta{\tool{C\kern-0.2exe\kern-0.5exT\kern-0.5exA}}

\newcommand\secref[1]{Section~\ref{sec:#1}}

\definecolor{funs}{rgb}{0.17,0.333,0.45}
\colorlet{vars}{black!30!orange}
\colorlet{urlblue}{blue!30!blue}
\newcommand\fs[1]{\texttt{\textcolor{funs}{#1}}}
\newcommand\fsop[1]{\mathbin{\fs{#1}}}
\newcommand\fsadd{\fsop{\char`+}}

\newcommand\0{\fs{0}}
\newcommand\s{\fs{s}}
\newcommand\f{\fs{f}}
\newcommand\g{\fs{g}}
\newcommand\h{\fs{h}}
\newcommand\var[1]{\mathit{\textcolor{vars}{#1}}}
\newcommand\x{\var{x}}
\newcommand\y{\var{y}}

\newcommand\interp[1]{[#1]}

\newcommand\extlinkicon{\faicon{%
  check-square-o%
}}

\newcommand\extlink[1]{%
  \href{#1}{\textcolor{urlblue}{\extlinkicon}}}

\newcommand\urlex[1]{\hfill\rlap{\extlink{#1}}}

\newcommand\urlexa{\urlex{%
  http://colo6-c703.uibk.ac.at/ttt2/web/?problem=(VAR\%20x\%20y)\%0A%
  (RULES\%0A\%20\%20\%20\%20\%2B(0\%2C\%20y)\%20-\%3E\%20y\%0A\%20\%2B(s(x)\%2C\%20y)\%20-\%3E\%20s(\%2B(x\%2C\%20y))\%0A)&strategy=kbo&template=\%2B\%20\%3E\%20s\%20\%3E\%200&template1=\%2B\%20\%3D\%20s\%20\%3D\%200\%20\%3D\%201&template2=1}}

\newcommand\urlexb{\urlex{%
  http://colo6-c703.uibk.ac.at/ttt2/web/?problem=(VAR\%20x\%20y)\%0A(RULES\%0A\%20\%20\%20\%20\%2B(0\%2C\%20y)\%20-\%3E\%20y\%0A\%20\%2B(s(x)\%2C\%20y)\%20-\%3E\%20s(\%2B(x\%2C\%20y))\%0A)&strategy=matrix2&template=0\%20\%3D\%200\%2C\%20s\%20\%3D\%20x0\%20\%2B\%201\%2C\%20\%2B\%20\%3D\%20\%5B1\%2C1\%3B0\%2C1\%5Dx0\%20\%2B\%20x1\%20\%2B\%20\%5B1\%3B0\%5D}}

\newcommand\urlexc{\urlex{%
  http://colo6-c703.uibk.ac.at/ttt2/web/?problem=(COMMENT\%20TRS_Standard\%2FSK90\%2F2.60.xml)\%0A(RULES\%20\%0A\%20\%20f(g(f(a)\%2Ch(a\%2Cf(a))))\%20-\%3E\%20f(h(g(f(a)\%2Ca)\%2Cg(f(a)\%2Cf(a))))\%20\%20\%20\%20\%20\%20\%0A)&strategy=lpo&template=NOT(AND(f\%20\%3E\%20g\%2C\%20f\%20\%3E\%20h))
}}

\newcommand\urlexd{\urlex{%
  http://colo6-c703.uibk.ac.at/ttt2/web/?problem=(COMMENT\%20TRS_Standard\%2FSK90\%2F4.11.xml)\%0A(VAR\%20x\%20y)\%0A(RULES\%20\%0A\%20\%20\%2B(x\%2C0)\%20-\%3E\%20x\%0A\%20\%20\%2B(x\%2Cs(y))\%20-\%3E\%20s(\%2B(x\%2Cy))\%0A\%20\%20\%2B(0\%2Cs(y))\%20-\%3E\%20s(y)\%0A\%20\%20s(\%2B(0\%2Cy))\%20-\%3E\%20s(y)\%20\%20\%20\%20\%20\%20\%0A)&strategy=poly&template=AND(\%2B\%20\%3D\%20_\%20\%2B\%202\%2C\%20OR(NOT(0\%20\%3D\%200)\%2C\%20s\%20\%3D\%20x0\%20\%2B\%201))
}}

\newcommand\urlexe{\urlex{%
  http://colo6-c703.uibk.ac.at/ttt2/web/?problem=(COMMENT\%20TRS_Standard\%2FDer95\%2F27.xml)\%0A(VAR\%20x\%20y)\%0A(RULES\%20\%0A\%20\%20h(f(x)\%2Cy)\%20-\%3E\%20f(g(x\%2Cy))\%0A\%20\%20g(x\%2Cy)\%20-\%3E\%20h(x\%2Cy)\%20\%20\%20\%20\%20\%20\%0A)&strategy=matrix3&template=f\%20\%3D\%20g\%20\%3D\%20h\%20\%3D\%20\%5B1\%2C_\%2C_\%3B0\%2C1\%2C_\%3B0\%2C0\%2C1\%5Dx0\%20\%2B\%20_\%2C\%20g\%20\%3D\%20h\%20\%3D\%20\%5B1\%2C_\%2C_\%3B0\%2C1\%2C_\%3B0\%2C0\%2C1\%5Dx1\%20\%2B\%20_
}}

\newcommand\texttildemid{\raisebox{.553ex}{\hbox{\texttildelow}}}

\lstdefinestyle{bashstyle}{%
  upquote=true,%
  basicstyle=\ttfamily\color{black},%
  basewidth={0.5em,0.45em},%
  literate=*%
    {\~}{{{\color{black}\texttildemid}}}1
}

\newcommand\sh[1]{\text{%
  \lstinline[style=bashstyle,%
    basicstyle=\ttfamily\color{black}]^#1^}}

\lstnewenvironment{bash}{%
  \lstset{style=bashstyle}
}{}

\title{\ttt2 with Termination Templates for Teaching%
\footnote{This work is supported by the Austrian Science Fund (FWF) project P27502.}}

\author{Jonas Schöpf}
\author{Christian Sternagel}
\affil{University of Innsbruck\\
  Innsbruck, Austria\\
  \texttt{\{jonas.schoepf|christian.sternagel\}@uibk.ac.at}}
\authorrunning{J. Schöpf and C. Sternagel} 

\Copyright{Jonas Schöpf and Christian Sternagel}

\EventEditors{Salvador Lucas}
\EventNoEds{1}
\EventLongTitle{Proceedings of the 16th International Workshop on Termination}
\EventShortTitle{WST 2018}
\EventAcronym{WST}
\EventYear{2018}
\EventDate{July 18--19, 2018}
\EventLocation{Oxford, UK}

\railnontermfont{\rmfamily\itshape}%
\railterm{tilde,underscore}%
\railalias{tilde}{\texttt{\char`~}}%
\railalias{underscore}{\texttt{\char`_}}%
\railparam{\setlength{\topsep}{5pt plus 1pt minus 2pt}}%

\begin{document}

\maketitle

\begin{abstract}
On the one hand, checking specific termination proofs by hand, say using a
particular collection of matrix interpretations, can be an arduous and
error-prone task.
On the other hand, automation of such checks would save time and help to
establish correctness of exam solutions, examples in lecture notes etc.
To this end, we introduce a template mechanism for the termination tool \ttt2
that allows us to restrict parameters of certain termination methods.
In the extreme, when all parameters are fixed, such restrictions result in
checks for specific proofs.
\end{abstract}

\section{Introduction}
\label{sec:intro}

Many of us are familiar with the following two (or at least similar) situations:
\begin{quote}\sl
Enthusiastically we call on our favorite termination tool in order to create an
example for a lecture or an exercise for an exam. But is it really necessary
that the weights of this KBO are higher than four? And wouldn't it be nicer if
the successor symbol was actually interpreted as the successor function in that
polynomial termination proof.

\smallskip
Hard pressed for time, you have to correct 20 term rewriting exams and each of
the students seems to have chosen different matrix interpretations in Example~2.
How will you manage before the deadline? Maybe you should just give each student
full points.
\end{quote}

As a more concrete example, consider the following well-known term rewrite
system (TRS) implementing addition on natural numbers:
\begin{xalignat*}{2}
\0 \fsadd \y &\to \y
&
\s(\x) \fsadd \y &\to \s(\x \fsadd \y)
\end{xalignat*}
If you ask \ttt2~\cite{KSZM2009} whether this TRS is terminating by KBO
(\sh{ttt2 -s 'kbo'}), you will get a \sh{YES} with $w_0 = 1$, weights~$w(\s) =
w(\0) = 1$ and $w({\fsadd}) = 0$, and precedence~${\fsadd} > \s \sim
\0$.\footnote{Where $\sim$ is ``don't care'' for strict and equivalence of
function symbols for quasi-precedences.}
In a lecture, you might want to restrict to the basic version
of KBO with total precedences and moreover it might be nicer for a presentation
if all function symbols had the same weight, say $1$.
Using our templates, this is now possible via
\begin{quote}
\sh{ttt2 -s 'kbo -prec "+ > s > 0" -w0 1 -weights "+ = s = 0 = 1"'}\urlexa
\end{quote}

An instance of the other kind of example we mention above would be the question
whether the following matrix interpretations prove termination of the addition
TRS (we can verify by \sh{ttt2 -s 'matrix -direct -dim 2'} that there is a
similar proof, but not using exactly the same interpretations):
\begin{xalignat*}{3}
\interp{\0} &= \left(\begin{matrix}0\\0\end{matrix}\right)
&
\interp{\s}(x_0) &= x_0 + \left(\begin{matrix}1\\1\end{matrix}\right)
&
\interp{{\fsadd}}(x_0,x_1) &=
\left(\begin{matrix}1&1\\0&1\end{matrix}\right)x_0 +
\left(\begin{matrix}1&0\\0&1\end{matrix}\right)x_1 +
\left(\begin{matrix}1\\0\end{matrix}\right)
\end{xalignat*}
With our template mechanism you can check the given proof by
\begin{quote}
\sh{ttt2 -s 'matrix -inters "0 = 0, s = x0 + 1, + = [1,1;0,1]x0 + x1 + [1;0]"'}
\urlexb
\end{quote}

The goal of our work is to successfully handle situations like the above.
More specifically, our contributions are as follows:
\begin{itemize}
\item
Based on a simple idea (\secref{idea}), we devised a template mechanism
(\secref{templates}) for four of the most prominent encoding-based termination
methods that allows us to employ \ttt2 as a ``proof checker.''

\item
Moreover, we extended \ttt2's web interface (\secref{web}, where also
{\scriptsize\extlink{}} is explained): on the one hand, to
support our template mechanism and on the other hand, by the possibility to
generate URLs that fix input TRSs and custom settings.
The latter is especially useful for lectures, where it provides a fast and easy
way to demonstrate examples.
\end{itemize}

\section{Main Idea}
\label{sec:idea}

Many of the termination methods that \ttt2 supports are based on SAT/SMT
encodings, notably the \emph{lexicographic path order} (LPO), the
\emph{Knuth-Bendix order} (KBO), \emph{polynomial interpretations} (PIs), and
\emph{matrix interpretations} (MIs).\footnote{There are others, like
\emph{arctic interpretations},
the \emph{(generalized) subterm criterion}, etc. But
for the moment templates are only supported by the four methods mentioned above.}

These methods have the following implementation detail in common: constraints on
their parameters (like a precedence for LPO or KBO and the maximal value of
matrix entries for MIs) are encoded into a SAT/SMT formula $\phi$ and then a
SAT/SMT-solver is used to obtain a concrete instance.
As a consequence such concrete instances of methods seem entirely random to the
casual user, which is not always desirable.

For example, we might only be interested in KBO proofs where a certain symbol
has least precedence, or we might want to interpret a unary function symbol
$\fs{s}$ as successor function without restricting the interpretations of other
symbols.
In the extreme, we already have a specific proof at hand (say using specific
matrix interpretations) and want to use a tool like \ttt2 to verify its
correctness.

In all of the above cases, an obvious solution is to modify the encoding~$\phi$ in
such a way that the desired constraints are fulfilled by every satisfying
assignment.  This is easily achieved by appending a formula $\chi$ that
represents these constraints, resulting in $\phi' = \phi \land \chi$.

Of course, it would be \emph{really} cumbersome if we had to write $\chi$ by
hand. Not to mention that we would have to know implementation details
concerning the encoding $\phi$, for example, which arithmetic variable
represents the precedence position of a given function symbol?

Instead we provide a method-specific template mechanism whose main work is to
parse and translate constraints that a user provides in form of a human-readable string.

\section{Templates for SAT/SMT-Based Methods}
\label{sec:templates}%

In the following we give an overview of the various templates that we provide
for the four encoding-based methods from the introduction.

\noindent
\begin{minipage}[t]{.48\textwidth}
\subparagraph{Lexicographic Path Orders}
The only parameter of an LPO (\sh{ttt2 -s 'lpo'}) is its precedence.  It can be
specified using the flag~\sh{-prec} that takes a \emph{prec} template as
argument.
A \emph{prec} template represents a (partial) precedence by listing its
constituent function symbols in decreasing order (separated by \texttt{>},
\texttt{=}, or \texttt{>=}), according to the syntax diagram on the right-hand side (as soon
as \texttt{=} or \texttt{>=} is used, we implicitly switch from
strict to quasi-precedences).
\end{minipage}
\hfill
\begin{minipage}[t]{.48\textwidth}
\begin{rail}
prec : fun (('>' | '=' | '>=') fun +)
;
\end{rail}
\end{minipage}

\subparagraph{Knuth-Bendix Orders}
In addition to a precedence,
KBO (\sh{ttt2 -s 'kbo'})
is also parameterized by weights, which can be
specified using the flags~\sh{-w0} (for the weight of variables)
and~\sh{-weights} that take a single weight and a \emph{weights} template,
respectively, as arguments.
Weights are natural numbers and can be specified for multiple function symbols
at once,
as depicted in
the following syntax diagram:
\begin{rail}
weights : (fun '=' *) fun ('=' | '<=' | '>=') weight
\end{rail}
If the last relation symbol of a \emph{weights} template is \texttt{<=} or
\texttt{>=}, then the specified weight is an upper and lower bound,
respectively, of all preceding function symbols of the same template.

\subparagraph{Linear Interpretations}
Linear interpretations can be specified using the flag~\sh{-inters} (supported
by PIs and MIs) that takes an \emph{inters} template as argument.
Such interpretations are sums of linear monomials that may either be an optional
coefficient followed by a variable, a constant part, or an underscore, according
to the syntax diagram:
\begin{rail}
inters : (fun '=' +) ((const? var | const | underscore) + '+')
;
\end{rail}
Here, underscores denote arbitrary (unrestricted) parts of an interpretation and
variables are associated to function arguments via their index (starting from
$0$). For example, given a binary function symbol $\f$, the template
\sh{2x0 + x1} fixes its interpretation to $\interp{\f}(\x,\y) = 2\x + \y$.

\subparagraph{Polynomial Interpretations}
Polynomial interpretations (\sh{ttt2 -s 'poly'}) are parameterized by linear
polynomials as interpretations for function symbols.
We obtain them by taking natural numbers for ``const'' in the \emph{inters}
template.

\bigskip

\noindent
\setlength{\fboxsep}{0pt}%
\begin{minipage}[t]{.42\textwidth}
\subparagraph{Matrix Interpretations}
For matrix interpretations (\sh{ttt2 -s 'matrix'}),
we instantiate ``const'' in the \emph{inters} template by matrices of natural
numbers as specified by the diagram on the right-hand side.%
\footnotemark
Moreover, we provide the shorthands~\sh{0} and~\sh{1} for the zero-vector and
one-vector, respectively.
\end{minipage}
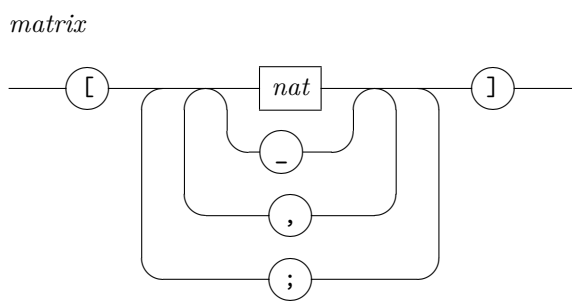
\hfill
\begin{minipage}[t]{.52\textwidth}
\begin{rail}
matrix : '[' (((nat | underscore) + ',') + ';') ']'
\end{rail}
\end{minipage}
\footnotetext{When the flag~\sh{-inters} is present, the matrix
dimension is read from the template, which usually means that \sh{-dim} is not
required anymore.}%

\subparagraph{Boolean Combinations of Constraints}
For all of the above templates (as atoms), we actually support arbitrary boolean
combinations of atomic constraints as specified below:
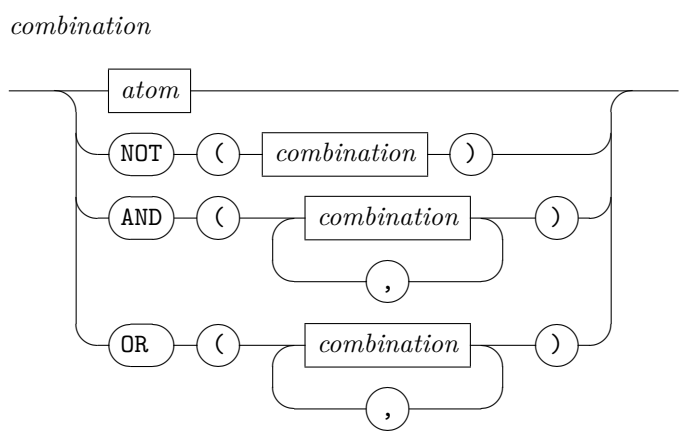
\begin{rail}
combination : atom
     | 'NOT' '(' combination ')'
     | 'AND' '(' (combination + ',') ')'
     | 'OR '  '(' (combination + ',') ')'
;
\end{rail}
As a common special case a comma-separated list of atoms is allowed at the
toplevel, which is then interpreted as logical conjunction of atomic templates.

\subparagraph{Examples}
We conclude this section by giving some example templates:
\begin{itemize}
\item
Let us first consider an LPO such that $\f$ is not at the same time bigger than
$\g$ and bigger than $\h$ in the precedence:
\sh{-prec "NOT(AND(f > g, f > h))"}
\urlexc

\item
Can we find (linear) polynomial interpretations such that the constant part of
$\interp{\fsadd}$ is $2$ and whenever $\0$ is interpreted as $0$, then $\s$
should also obtain its natural interpretation?
\\
\sh{-inters "AND(+ = _ + 2, OR(NOT(0 = 0), s = x0 + 1))"}
\urlexd

\item
How about matrix interpretations with upper triangular matrices of dimension
three that have only ones on their diagonals, for unary $\f$ and binary $\g$ and
$\h$?
\\
\sh{-inters "f=g=h=[1,_,_;0,1,_;0,0,1]x0+_, g=h=[1,_,_;0,1,_;0,0,1]x1+_"}
\urlexe
\end{itemize}

\section{Teaching with the Web-Interface}
\label{sec:web}

It is the experience of the second author that for teaching, while live
demonstrations of tools are often appreciated by students, properly setting up
such examples often takes way too much time. To remedy this situation at least
for \ttt2, we integrated a mechanism into its web
interface\footnote{\url{http://colo6-c703.uibk.ac.at/ttt2/web/}} that allows a user
to store the current configuration (including the input TRS) into the query part
of the web interface URL.

Now, we can easily copy such a URL and turn it, for example, into a PDF hyperlink
on slides or in a paper. Incidentally, this is exactly the purpose of the
{\scriptsize\extlink{}} symbols in this paper. The first one, for example, is
generated by the following (incomplete) \LaTeX{} code:
\begin{verbatim}
  \href{http://colo6-c703.uibk.ac.at/ttt2/web/?problem=(VAR\%20x\%20y)...
\end{verbatim}

Moreover, our termination templates are accessible through
dedicated input fields of the web interface. Thus, a user does not have to type
(and know) the corresponding flags.

\section{Conclusion and Future Work}
\label{sec:concl}

We presented the extension of \ttt2 by a template mechanism for four of the
most common termination techniques that are taught in classes. These templates
allow us to narrow down the search space such that in the extreme, we are left
with a specific instance of a termination technique. In this way, \ttt2 can be
used as a ``proof checker'' for termination proofs (if you want to have near
absolute certainty that some termination proof is correct, you should of course
validate \ttt2's output by a formally verified certifier like \ceta).
Furthermore, we demonstrated a simple but useful addition to \ttt2's web
interface to generate URLs that make the input TRS and current configuration
persistent.
%

We leave templates for methods like
\emph{arctic interpretations}~\cite{KW2008},
the \emph{(generalized) subterm criterion}~\cite{S2016},
and \emph{finding loops}~\cite{ZSHM2010} as future work.
An orthogonal issue that requires further investigation is: which extensions to
the current templates would be most useful for teaching?

Some similar options to our termination templates were available through a GUI in an old
version of \aprove~\cite[Section~2]{A_RTA2004}. Unfortunately, this GUI was abandoned in
later versions.

\bibliographystyle{plain}
\bibliography{references}

\end{document}